\documentstyle[12pt]{article}

\begin{document}

\baselineskip=18pt plus 0.2pt minus 0.1pt
\parskip = 6pt plus 2pt minus 1pt

%%%%%%%%%%%%%%%%%%%%%% PRIVATE MACROS %%%%%%%%%%%%%%%%%%%%%%%%%%%%%%%%%%%%%
\catcode`\@=11

\newif\iffigureexists
\newif\ifepsfloaded
\openin 1 epsf.sty
\ifeof 1 \epsfloadedfalse \else \epsfloadedtrue \fi
\closein 1
\ifepsfloaded
    \input epsf.sty
\else
    \immediate\write20{>Warning:
         No epsf.sty --- cannot embed Figures!!}
\fi
\def\checkex#1 {\relax
    \ifepsfloaded \openin 1 #1
        \ifeof 1 \figureexistsfalse
        \else \figureexiststrue
        \fi \closein 1
    \else \figureexistsfalse
    \fi }

\def\epsfhako#1#2#3#4#5#6{
%% #1:eps-file-name #2:htb etc #3:epsfxsize #4:Caption
%% #5:label #6: \vspace etc between Fig and Caption.
\checkex{#1}
\iffigureexists
    \begin{figure}[#2]
    \epsfxsize=#3
    \centerline{\epsffile{#1}}
    {#6}
    \caption{#4}
    \label{#5}
    \end{figure}
\else
    \begin{figure}[#2]
    \caption{#4}
    \label{#5}
    \end{figure}
    \immediate\write20{>Warning:
         Cannot embed a Figure (#1)!!}
\fi
}

\ifepsfloaded
\checkex{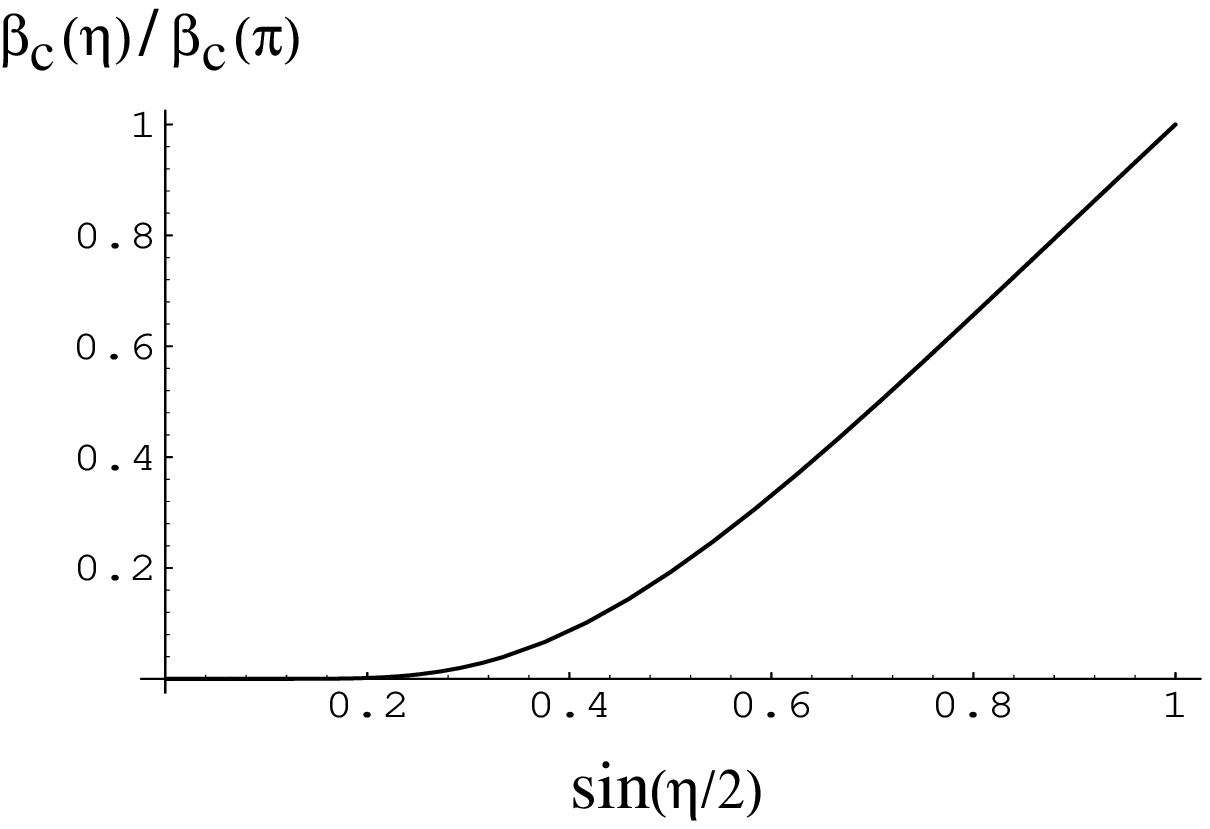}
    \iffigureexists \else
    \immediate\write20{>EPS files for Figs. 2 and 3 are packed
     in a uuecoded compressed tar file}
    \immediate\write20{>appended to this LaTeX file.} 
    \immediate\write20{>You should unpack them and LaTeX again!!}
    \fi
\fi

\renewcommand{\thefootnote}{\fnsymbol{footnote}}
\renewcommand{\theequation}{\thesection.\arabic{equation}}
\newcommand{\reseteqnum}{\setcounter{equation}{0}}
\newcommand{\eq}[1]{eq.\ (\ref{#1})}
\newcommand{\tr}{{\rm tr}}
\newcommand{\Tr}{{\rm Tr}}
\newcommand{\bm}[1]{\mbox{\boldmath $#1$}}
\newcommand{\ovl}[1]{\overline{#1}}
\newcommand{\wt}[1]{\widetilde{#1}}
\newcommand{\vev}[1]{\langle #1 \rangle}
\newcommand{\bvev}[1]{\vev{#1}_{\!\lower2pt\hbox{$\scriptstyle\beta$}}}
\newcommand{\dB}{\delta_{\rm B}}
\newcommand{\dA}{\delta_{\rm A}}
\newcommand{\e}{\epsilon}
\newcommand{\p}{\partial}
\newcommand{\bc}{\overline{c}}
\newcommand{\bB}{\overline{B}}
\newcommand{\QB}{Q_{\rm B}}
\newcommand{\wtQB}{\widetilde{Q}_{\rm B}}
\newcommand{\olQB}{\overline{Q}_{\rm B}}
\newcommand{\calL}{{\cal L}}
\newcommand{\calD}{{\cal D}}
\newcommand{\calO}{{\cal O}}
\newcommand{\calZ}{{\cal Z}}
\newcommand{\Ngh}{N_{\rm gh}}
\newcommand{\Pphys}{{\cal P}_{\rm \! phys}}
\newcommand{\1}{1\kern-5pt 1}
\newcommand{\Half}{\frac{1}{2}}
\newcommand{\vphi}{\mbox{\boldmath $\varphi$}}
\newcommand{\svphi}{\mbox{\boldmath $\scriptstyle\varphi$}}
\newcommand{\vj}{\mbox{\boldmath $j$}}
\newcommand{\vL}{\mbox{\boldmath $L$}}
\newcommand{\svL}{\mbox{\boldmath $\scriptstyle L$}}
\newcommand{\veta}{\mbox{\boldmath $\eta$}}
\newcommand{\sveta}{\mbox{\boldmath $\scriptstyle \eta$}}
\newcommand{\bra}[1]{\left\langle #1\right|}
\newcommand{\ket}[1]{\left| #1\right\rangle}
\newcommand{\braket}[2]{\vev{#1 | #2}}
\newcommand{\brabeta}{\bra{0(\beta)}}
\newcommand{\ketbeta}{\ket{0(\beta)}}
\newcommand{\OSp}{O\!Sp}
\newcommand{\NLSM}{NL$\sigma$M}
\newcommand{\lrp}{\mathop{\partial}\limits^{\leftrightarrow}}
\newcommand{\sech}{{\rm sech}}
\renewcommand{\Box}{\kern1pt\vbox{\hrule height .6pt
            \hbox{\vrule width .6pt\hskip 3pt
            \vbox{\vskip 8pt}\hskip 5pt\vrule width .6pt}
            \hrule height .6pt}\kern1pt}
%%%
\def\gtlt{\mathrel{\mathpalette\@verlt>}}
\def\@verlt#1#2{\vcenter{\offinterlineskip
        \ialign{$\m@th#1\hfil##\hfil$\crcr#2\crcr<\crcr } }}
%%%

\catcode`\@=12

%%%%%%%%%%%%%%%%%%%%%%%%%%%%
\newcommand{\figC}{
\begin{figure}[htbp]
\begin{center}
        \begin{picture}(300,170)
                \put(20,100){\line(1,0){260}}
                \put(150,0){\line(0,1){160}}
                \put(40,98){\line(0,1){4}}
                \put(140,170){Im $t$}
                \put(285,95){Re $t$}
                \put(40,110){$-T$}
                \put(260,110){$+T$}
                \put(40,65){$-T\!-i\frac{\beta}{2}$}
                \put(260,45){$+T\!-i\frac{\beta}{2}$}
                \put(40,0){$-T\!-i\beta$}
                \put(130,105){$C_1$}
                \put(130,60){$C_2$}
                \put(265,70){$C_3$}
                \put(45,30){$C_4$}
        \thicklines
                \put(40,100){\vector(1,0){55}}
                \put(150,100){\vector(1,0){55}}
                \put(260,55){\vector(-1,0){55}}
                \put(150,55){\vector(-1,0){55}}
                \put(260,100){\vector(0,-1){22}}
                \put(40,55){\vector(0,-1){22}}
                \put(40,100){\line(1,0){220}}
                \put(260,100){\line(0,-1){45}}
                \put(40,55){\line(1,0){220}}
                \put(40,55){\line(0,-1){45}}
        \end{picture}
\end{center}
\caption{The time contour $C$.}
\label{fig:C}
\end{figure}
}
%%%%%%%%%%%%%%%%%%%%%%%%%%%%

\begin{titlepage}
\title{
\hfill
\parbox{4cm}{\normalsize KUNS-1260\\HE(TH)~94/06\\hep-th/9405145}\\
\vspace{1cm}
Finite Temperature Deconfining Transition\\ in the BRST Formalism
}
\author{Hiroyuki Hata\thanks{e-mail address:
\tt hata@gauge.scphys.kyoto-u.ac.jp, hata@jpnyitp.bitnet}
{\,}\thanks{Supported in part by Grant-in-Aid for Scientific Research
from Ministry of Education, Science and Culture (\# 06640390).}
{} and Yusuke Taniguchi\thanks{e-mail address:
\tt tanigchi@gauge.scphys.kyoto-u.ac.jp}
{\,}\thanks{JSPS Research Fellow.
Supported in part by Grant-in-Aid for Scientific Research
from Ministry of Education, Science and Culture (\# 3113).}
\\
{\normalsize\em Department of Physics, Kyoto University}\\
{\normalsize\em Kyoto 606-01, Japan}}
\date{\normalsize May, 1994}
\maketitle
\thispagestyle{empty}

\begin{abstract}
\normalsize
We present a toy model study of the high temperature deconfining
transition in Yang-Mills theory as a breakdown of the confinement
condition proposed by Kugo and Ojima.
Our toy model is a kind of topological field theory obtained from the
Yang-Mills theory by taking the limit of vanishing gauge
coupling constant $g_{\rm YM}\to 0$, and therefore the gauge field
$A_\mu$ is constrained to the pure-gauge configuration
$A_\mu=g^{\dagger}\partial_\mu g$.
At zero temperature this model has been known to satisfy the
confinement condition of Kugo and Ojima
which requires the absence of the massless Nambu-Goldstone-like mode
coupled to the BRST-exact color current.
In the finite temperature case based on the real-time formalism, our
model in 3+1 dimensions is reduced, by the Parisi-Sourlas mechanism,
to the ``sum'' of chiral models in 1+1 dimensions with various
boundary conditions of the group element $g(t,x)$ at the ends of
the time contour. We analyze the effective potential of the $SU(2)$
model and find that the deconfining transition in fact occurs due to
the contribution of the sectors with non-periodic boundary conditions.
\end{abstract}
\end{titlepage}

\newpage
\section{Introduction}
\reseteqnum

Deconfining transition in high temperature QCD has been a subject of
active researches for both theoretical and phenomenological interests.
This transition may be stated as follows: below a certain critical
temperature $\beta_c^{-1}$ we cannot observe color non-singlet
excitations, while above $\beta_c^{-1}$ colored excitations such as
quarks are allowed as physical ones.
Widely adopted as a criterion of (de)confinement at finite temperature
is the expectation value $\bvev{P}$ of the Polyakov loop (thermal
Wilson loop) operator
$P(\bm{x}) = \tr\,{\rm P}\exp\left(
\int_0^\beta\! d\tau A_4(\tau,\bm{x})\right)$
which measures the free-energy of a quark
put in the system as an external source \cite{Polyakov,Susskind}.
However, the Polyakov loop cannot be used as a criterion of
confinement in a system with dynamical (quantized) quark fields.
Moreover, $\bvev{P}$ tells nothing about the (de)confinement of other
color carrying fields, for example, the gluon.
In this paper we shall study the deconfining transition
on the basis of another confinement criterion proposed by Kugo and
Ojima (KO) \cite{KO}. In contrast with the Polyakov loop,
the confinement mechanism of KO treats directly the confinement of
quantized colored fields (particles) of any kind.

The KO confinement criterion is based on the BRST quantized Yang-Mills
theory \cite{KO} described by the lagrangian\footnote{
We restrict the gauge group to $SU(N)$.
The field variables $\phi=A_\mu,c,\bc,B$ are Lie algebra valued
and are expressed as $\phi=\sum_{a=1}^{N^2-1}\phi^a t^a$ in terms of
Hermitian fields $\phi^a$ and the (anti-Hermitian) basis $t^a$
with the normalization $\tr(t^at^b)=-(1/2)\delta^{ab}$.}
\begin{equation}
{\cal L}_{\rm YM} = \frac{1}{2g_{\rm YM}^2}\tr F_{\mu\nu}^2
+ {\cal L}_{\rm matter} -i \dB G ,
\label{eq:YMLagrangian}
\end{equation}
where $\dB$ is the BRST transformation defined as usual by
\begin{equation}
\dB A_\mu=D_\mu c\equiv \p_\mu c + [A_\mu,c],\quad
\dB c= -\Half\{c,c\},\quad
\dB\bc=iB,\quad
\dB B=0.
\label{eq:dB}
\end{equation}
The last term of \eq{eq:YMLagrangian} gives the gauge-fixing
and the corresponding ghost terms.
The key quantity in the KO confinement mechanism at zero temperature is
the BRST-exact conserved color current $N_\mu$,
\begin{equation}
N_\mu = -i\dB K_\mu = \{\QB,K_\mu\} ,
\label{eq:N}
\end{equation}
where $\QB$ is the BRST charge, and $K_\mu$ is obtained from
$\calL_{\rm YM}$ of \eq{eq:YMLagrangian} by making a local gauge
transformation $\delta_\e$, $\delta_\e A_\mu=D_\mu\e$ and
$\delta_\e\phi=[\phi,\e]$ for $\phi=c,\bc,B$, as
\begin{equation}
\delta_\e \calL_{\rm YM}=
-i\delta_\e \dB G= -i \dB K^a_\mu\cdot\p_\mu \e^a .
\label{eq:deL}
\end{equation}
Note that $\delta_\e\calL_{\rm YM}$ has its contribution only from the
gauge-term $\dB G$, and the last expression is because we have assumed
that the gauge-fixing function $G$ preserves the global color rotation
symmetry. In the Feynman-type gauge with
$G=\tr\left[\bc\left(\p_\mu A_\mu - \alpha B\right)\right]$,
we have $K_\mu=D_\mu\bc$. The ordinary Noether color current
$J^a_\mu=\ovl{q}\gamma_\mu t^a q+\ldots$
containing the matter fields is related to $N_\mu$ by
$J_\mu=N_\mu-(1/g_{\rm YM}^2)\p^\nu F_{\nu\mu}$ using the equation of
motion. Therefore both $N_\mu$ and $J_\mu$ generate the same global
color rotation.

The KO confinement condition requires that the BRST-exact color
current $N_\mu$ (\ref{eq:N}) contains no (Nambu-Goldstone-like)
massless one-particle component. If this condition is satisfied, then
the integration $\int\! d^3xN_0$ has a well-defined meaning and hence
the color charge $Q^a$ can be written in a BRST-exact form,
$Q^a=\left\{\QB, \int\! d^3x K^a_0\right\}$, which implies color
confinement in the sense that any color non-singlet asymptotic state
is necessarily BRST unphysical and unobservable.\footnote{
Detailed explanation of the KO confinement mechanism at zero
temperature is found in the original paper \cite{KO}
(see also ref.\ \cite{HataNiigata} for a brief explanation).
}

In order to extend the KO confinement mechanism to the finite
temperature case, we shall recapitulate the elements of finite
temperature gauge theory.
First, in statistical mechanics of gauge theories in the BRST
formalism, the statistical average must be taken only over physical
states.
Let $\Pphys$ be the projection operator to the subspace of physical
states ${\cal H}_{\rm phys}={\rm Ker}\,\QB/{\rm Im}\,\QB$.
Then we have the following useful identity \cite{FiniteT-HK} for the
thermal expectation value $\bvev{\calO}$ of a BRST invariant operator
$\calO$ satisfying $[\QB,\calO]=0$:
\begin{equation}
\bvev{\calO}\equiv\Tr\left(\Pphys e^{-\beta H}\calO\right)/\calZ(\beta) =
\Tr\left(e^{-\beta H + i\pi\Ngh}\calO\right)/\calZ(\beta) ,
\label{eq:HKformula}
\end{equation}
where $\Ngh$ is the ghost number operator, and $\Tr$ means the trace
operation over all (physical as well as unphysical) states:
\begin{equation}
\Tr\,\calO \equiv \sum_{k,l}\bra{k}\calO\ket{l}\eta^{-1}_{lk}
\quad \biggl(\eta_{kl}=\braket{k}{l},
\quad \sum_{l}\eta_{kl}\eta^{-1}_{lm}=\delta_{km}\biggr) .
\label{eq:Tr}
\end{equation}
The partition function $\calZ(\beta)$ itself is also given by
$\calZ(\beta)\equiv\Tr\left(\Pphys e^{-\beta H}\right)=
\Tr\left(e^{-\beta H + i\pi\Ngh}\right)$.
Eq.\ (\ref{eq:HKformula}) is a consequence of the formula $\1=\Pphys
+ \{\QB,{}^\exists R\}$ where $\1$ is the identity operator.
We adopt the last expression of \eq{eq:HKformula} with statistical
weight $e^{-\beta H + i\pi\Ngh}$ as the definition of $\bvev{\calO}$
even when $\calO$ is not a BRST invariant quantity.

A framework of finite temperature field theory which is suitable for
discussing the KO mechanism is the real-time formalism where we can
treat fields with ordinary time variable $t$ ($-\infty<t<\infty$).
There are two formulations of the real-time formalism. One is the
path-integral formalism \cite{NS}, and the other is the operator
formalism called thermo field dynamics (TFD) \cite{TakaUme,UMT}.
The path-integral formalism is convenient for concrete calculations,
while TFD is necessary to generalize the KO confinement mechanism,
which has been formulated in the BRST operator formalism, to the
finite temperature case. In this paper we shall use both of the two
formalisms regarding them as equivalent. We use the same symbol for
both an operator and the corresponding path-integration variable.

\figC

First the path-integral formalism applied to a gauge theory is as
follows. We consider thermal Green's functions generated by
\begin{equation}
Z[j]=\Tr\left\{e^{-\beta H+i\pi\Ngh}{\rm T}_C\!
\left[\exp i\int_C j\phi\right]\right\} ,
\label{eq:Zj-op}
\end{equation}
where we have used the abbreviation
$\int_C A\equiv\int_C d\tau\int d^3x A(\tau,\bm{x})$ with the contour
$C$ in the complex time-plane depicted in fig.\ \ref{fig:C},
and $T_C$ denotes the ordering along the contour $C$.
$\phi$ and $j$ represents a generic field operator in the system and
the corresponding source.
The generating functional $Z[j]$ (\ref{eq:Zj-op}) has a path-integral
expression \cite{NS}:
\begin{equation}
Z[j]=\int\limits_{\rm periodic}\!\!\!\calD\phi\,\exp\left\{
i\int_C\left(\calL_{\rm YM}(\phi) + j\phi\right)\right\} ,
\label{eq:Zj-pi}
\end{equation}
where the functional integration
$\calD\phi\equiv\calD A_\mu\,\calD c\,\calD\bc\,\calD B$
should be done with {\em periodic} boundary condition
$\phi(-T\!-i\beta)=\phi(-T)$ for all the fields $\phi=A_\mu,c,\bc,B$.
In particular, the effect of the factor $e^{i\pi\Ngh}$ in
\eq{eq:Zj-op} is to turn the boundary condition of the fermionic
fields $c$ and $\bc$ to the periodic one.
Since the contour $C$ of fig.\ \ref{fig:C} contains the
horizontal (real-time) segments $C_1$ and $C_2$ of infinite length
(we take the limit $T\to\infty$), we can consider thermal Green's
functions with ordinary real time arguments in contrast to the
imaginary-time formalism where the contour $C$ is simply a straight
vertical line $[0,-i\beta]$.

Thermo field dynamics (TFD) \cite{TakaUme,UMT} is the operator
formalism which reproduces the thermal average of \eq{eq:HKformula} as
an expectation value with respect to ``temperature dependent vacuum''
$\ketbeta$:
\begin{equation}
\bvev{\calO}=\brabeta\calO\ketbeta .
\label{eq:vev0beta}
\end{equation}
Application of TFD to gauge theories is given in ref.\ \cite{Ojima}.
Corresponding to two horizontal segments $C_1$ and $C_2$ of the
contour $C$ in fig.\ \ref{fig:C}, we have to double the fields in TFD
as compared to the theory at zero temperature.
The fields and states corresponding to $C_1$ is denoted as before by
$\phi(t,\bm{x})$ and $\ket{k}$, and those corresponding to $C_2$ is
denoted with tilde; $\wt{\phi}(t,\bm{x})$ and $\wt{\ket{k}}$.
Then $\ketbeta$ is given as
\begin{equation}
\ketbeta=\calZ(\beta)^{-1/2}\sum_{k,l}\exp\left(
-\Half\beta H +\frac{i\pi}{2}\Ngh\right)\ket{k}\otimes\wt{\ket{l}}
\,\eta^{-1}_{kl} .
\label{eq:0beta}
\end{equation}
One can easily see that \eq{eq:vev0beta} holds for this $\ketbeta$.
The lagrangian $\ovl{\calL}_{\rm YM}$ describing TFD is
\begin{equation}
\ovl{\calL}_{\rm YM}=\calL_{\rm YM} - \wt{\calL}_{\rm YM} ,
\label{eq:LinTFD}
\end{equation}
where $\wt{\calL}_{\rm YM}$ is the lagrangian for the tilde fields
$\wt{\phi}$.
The physical states in TFD are specified by $\olQB\ket{\mbox{phys}}=0$
using the BRST charge of the whole system $\olQB \equiv \QB - \wtQB$.
See refs.\ \cite{Ojima} for precise definitions.

Now we are ready to generalize the KO (de)confinement criterion to the
finite temperature Yang-Mills theory.
In the finite temperature case based on TFD, the relevant BRST-exact
color current has contributions from both the non-tilde and tilde
fields (cf. \eq{eq:LinTFD}):
\begin{equation}
\ovl{N}_\mu = N_\mu - \wt{N}_\mu = \{\olQB,\ovl{K}_\mu\} ,
\label{eq:olN}
\end{equation}
where $\ovl{K}_\mu\equiv K_\mu - \wt{K}_\mu$.
Note that the vacuum $\ketbeta$ (\ref{eq:0beta}) is not invariant
under a separate color rotation on non-tilde or tilde fields generated
by $N_\mu$ or $\wt{N}_\mu$.
Using this $\ovl{N}_\mu$, the KO confinement mechanism is generalized
to the finite temperature Yang-Mills system as follows.
{\em
If the BRST-exact color current $\ovl{N}_\mu$ (\ref{eq:olN})
contains no massless one-particle component, it implies color
confinement: the system contains no
colored excitations (quasi-particles) as physical ones.\footnote{
We are assuming that the asymptotic field like analysis applies also
to TFD.}
}

In order to study the deconfining transition, we have to first prepare
a situation where the confinement is realized at zero
temperature ($\beta=\infty$). Although the KO confinement
criterion has not been shown to hold in real QCD at $\beta=\infty$,
we have a toy model of four-dimensional Yang-Mills system where the
confinement condition of KO is known to be satisfied at
$\beta=\infty$. This toy model is obtained from the real Yang-Mills
system (\ref{eq:YMLagrangian}) by taking the limit of vanishing gauge
coupling constant; $g_{\rm YM}\to 0$.
In this limit, the field strength term,
$(1/g_{\rm YM}^2)\tr F_{\mu\nu}^2$, of \eq{eq:YMLagrangian} forces the
gauge field $A_\mu$ constrained to the {\em pure-gauge} configuration,
$A_\mu(x)=g^\dagger(x)\p_\mu g(x)$, and hence the system is reduced to
something like a topological field theory described by the BRST-exact
lagrangian  $-i\dB G[A_\mu=g^\dagger\p_\mu g]$ \cite{PGM-H,PGM-HK}.
We call this toy model the pure-gauge model.

Although the pure-gauge model contains no physical degrees of
freedom, the KO confinement criterion is still a non-trivial dynamical
problem. In naive perturbation theory of both the real Yang-Mills
theory and the pure-gauge model at zero temperature, we have a
massless pole coupled to $N_\mu$. This is revealed by the Green's
function,
\begin{equation}
\int\!\frac{d^4p}{(2\pi)^4}e^{ip\cdot x}\vev{{\rm T}N^a_\mu(x)
A^b_\nu(0)}_{\beta=\infty}
\sim \delta^{ab}\frac{p_\mu p_\nu}{p^2} .
\label{eq:NA}
\end{equation}
However, in the case of pure-gauge model with a special gauge called
$OSp(4/2)$ symmetric gauge, we can show the massless pole in
(\ref{eq:NA}) is in fact missing. This is because the pure-gauge model
in 3+1 dimensions is shown to be ``equivalent'' by the Parisi-Sourlas
mechanism \cite{ParisiSourlas} to the chiral model in 1+1 dimensions
where the vacuum is realized in disordered phase with a mass gap
\cite{PolyakovWiegmann}.
It is expected that, in the real Yang-Mills theory also, the KO
confinement condition (at $\beta=\infty$) holds due to large gauge
field fluctuation in the direction of local gauge
transformation\rlap.\footnote{
In fact, the color confinement by the KO mechanism is interpretable as
a consequence of the {\em restoration} of local gauge symmetry with
transformation parameter $\e(x)\sim a_\mu x^\mu$ ($a_\mu$: const.)
and hence $\delta_\e A_\mu \sim a_\mu$ \cite{Hata-restoration}.
}

Having completed the preparation, let us turn to the explanation of
our analysis of finite temperature case carried out in this paper.
Recalling that the pure-gauge model is obtained as the $g_{\rm YM}\to
0$ limit of Yang-Mills theory (\ref{eq:YMLagrangian}), we consider
the two-point function
\begin{equation}
\lim_{g_{\rm YM}\to 0}
\int\!\frac{d^4p}{(2\pi)^4}e^{ip\cdot x}\vev{{\rm T}\ovl{N}^a_\mu(x)
A^b_\nu(0)}_{\beta} ,
\label{eq:limNA}
\end{equation}
in the limit $g_{\rm YM}\to 0$.
(In eq.\ (\ref{eq:limNA}), T denotes the time-ordering in TFD and
the ${\rm T}_C$-ordering in the path-integral formalism.)
We know that (\ref{eq:limNA}) is free
from massless poles at zero temperature $\beta=\infty$, and study
whether a massless pole is generated at high temperature
$\beta<\beta_c$ with some critical $\beta_c$.
For this purpose we observe that in the path-integral formalism
(\ref{eq:Zj-pi}) the $A_\mu$-integration is reduced in the
$g_{\rm YM}\to 0$ limit to the $g$-integration
($A_\mu=g^\dagger\p_\mu g$) and, in addition, that, although
$A_\mu=g^\dagger\p_\mu g$ has to satisfy the periodic boundary
condition $A_\mu(-T\!\!-i\beta,\bm{x})=A_\mu(-T,\bm{x})$, the group
element $g(t,\bm{x})$ need not be strictly periodic. In fact, the
boundary conditions using the constant $SU(N)$ left-transformations,
$g(-T\!\!-i\beta,\bm{x})=h\cdot g(-T,\bm{x})$,
are allowed ones which respect the periodicity of $g^\dagger\p_\mu g$.
Adopting the $\OSp(4/2)$ symmetric gauge of ref.\ \cite{PGM-HK} we
find that the two-point function (\ref{eq:limNA}) in 3+1 dimensions is
equal to the {\em average} of the Green's function
$\int\!d^2p/(2\pi)^2 e^{ip\cdot x}\vev{{\rm T}\ovl{A}^a_\mu(x)
A^b_\nu(0)}_{\beta}$ over the boundary conditions of $g(t,x)$
in the 1+1 dimensional chiral model.
This implies that color confinement by the KO mechanism breaks
down if the corresponding chiral models in 1+1 dimensions are realized
in the ordered Nambu-Goldstone phase for a finite range of the
boundary conditions.

Since we do not have a systematic non-perturbative methods to analyze
finite temperature chiral model in the real-time formalism with
unusual boundary conditions, we shall carry out the analysis of the
effective potential of the $O(4)$ non-linear $\sigma$-model (which is
equivalent to the $SU(2)$ chiral model) obtained by the large-$N$ like
method.
Our analysis shows that there is indeed a desired deconfining
transition at $1/\beta\sim m$ ($m$: mass gap at $\beta=\infty$).
The transition occurs because the infrared singularity which realized
the disordered phase at zero temperature is weakened by the
non-periodic boundary conditions.

The rest of this paper is organized as follows.
In Sec.\ 2, we introduce the finite temperature pure-gauge model in
the real-time formalism as the $g_{\rm YM}\to 0$ limit of the
Yang-Mills theory, and explain the Parisi-Sourlas reduction to the
two-dimensional chiral model. In Sec.\ 3, we carry out the analysis of
the $O(4)$ non-linear $\sigma$-model with generalized boundary
conditions. The final section (Sec.\ 4) is devoted to summary and
discussion.

\section{The pure-gauge model}
\reseteqnum

As stated in the Introduction we shall consider the $g_{\rm YM}\to 0$
limit in finite temperature Yang-Mills system in the real-time formalism
described by the path-integral (\ref{eq:Zj-pi}).
In this limit the gauge field $A_\mu$ is restricted to the pure-gauge
configuration,
\begin{equation}
A_\mu(x)=g^\dagger(x)\p_\mu g(x)\quad\mbox{with}\quad g(x)\in SU(N) ,
\label{eq:pure-gauge}
\end{equation}
due to the $(1/g_{\rm YM}^2)\tr F_{\mu\nu}^2$ term in
${\cal L}_{\rm YM}$ (\ref{eq:YMLagrangian}), and the system is reduced
to the pure-gauge model (PGM) with dynamical variables $(g,c,\bc,B)$
just as in the zero-temperature case \cite{PGM-H,PGM-HK}.
What is particular to the finite temperature case is the boundary
condition of $g(t,\bm{x})$. Although $A_\mu=g^\dagger\p_\mu g$ has
to satisfy the periodic boundary condition,
\begin{equation}
A_\mu(-T\!-i\beta,\bm{x})=A_\mu(-T,\bm{x}) ,
\label{eq:PBofA}
\end{equation}
in \eq{eq:Zj-pi}, the group element $g(t,\bm{x})$ need not be strictly
periodic. Boundary conditions related by a constant $SU(N)$
left-transformation,
\begin{equation}
B_h : g(-T\!-i\beta,\bm{x}) = h\!\cdot\! g(-T,\bm{x})
\quad \left(h\in SU(N)\right),
\label{eq:Bh}
\end{equation}
are allowed ones which respect the periodicity of
$A_\mu=g^\dagger\p_\mu g$.

Therefore it is natural to assume that in the limit $g_{\rm YM}\to 0$
the $A_\mu$-integration with periodic boundary condition is reduced to
the $g$-integration using all the boundary conditions $B_h$; namely we
have to integrate over the boundary condition parameter $h$:\footnote{
We consider for simplicity a system without quark fields.
The quark fields do not decouple from the $(g,c,\bc,B)$ system if we
impose a non-periodic boundary condition on $g$.
}
\begin{equation}
\int\limits_{\rm periodic}\kern-10pt\calD A_\mu
\quad\to\quad
\int\! dh \int\nolimits_{B_h}\!\calD g \ ,
\label{eq:Dg}
\end{equation}
where $\int dh$ denotes the Haar measure of $SU(N)$.
Then the expectation value of an operator $\calO$ is reduced in the
$g_{\rm YM}\to 0$ limit to
\begin{equation}
\bvev{\calO}\to
\int\! dh\int_{B_h}\!\calD g\calD c\calD\bc\calD\! B\,
\calO \exp\left(i\int_C \calL_{\rm PGM}\right)
\Bigg/\int\! dh\,\calZ_h \ ,
\label{eq:vevinPGM1}
\end{equation}
where $\calL_{\rm PGM}$ is the lagrangian of PGM,
\begin{equation}
\calL_{\rm PGM}=-i\dB G\vert_{A_\mu=g^\dagger\p_\mu g} \ ,
\label{eq:LPGM}
\end{equation}
and $\calZ_h$ is the partition function of the PGM with the
boundary condition $B_h$.
The BRST transformation on $g$ is given by $\dB g=gc$.
The partition function $\calZ_h$ is shown to be equal to $1$ for any
$h$ by reversing the manipulation of eq.\ (\ref{eq:HKformula}):
\begin{equation}
\calZ_h=\Tr\left(U_h e^{-\beta H + i\pi \Ngh}\right)
= \Tr\left(U_h\,\Pphys e^{-\beta H}\right)=1 ,
\label{eq:Zk}
\end{equation}
where $U_h$ is the operator of the left-transformation by $h$,
$U_h^{-1}g U_h = h\!\cdot\! g$.
In eq.\ (\ref{eq:Zk}) use has been made of the property
$[\QB,U_h]=0$, and the facts that the vacuum $\ket{0}$ is the only
physical state in the PGM (at zero temperature) and that
$U_h\!\ket{0}=\ket{0}$ for any $h$. Therefore the expectation value
$\bvev{\calO}$ in the limit $g_{\rm YM}\to 0$ is expressed as the
{\em average\/} over $h$ of the expectation values
$\vev{\calO}_{\beta,h}$ in the PGM with the boundary condition $B_h$:
\begin{equation}
\bvev{\calO}\to
\int\! dh\frac{1}{\calZ_h}
\int_{B_h}\!\calD g\calD c\calD\bc\calD\! B\,
\calO \exp\left(i\int_C \calL_{\rm PGM}\right)
\equiv \int\! dh\,\vev{\calO}_{\beta,h} .
\label{eq:vevinPGM2}
\end{equation}
The integrations over $(c,\bc,B)$ should be done using periodic
boundary condition.

The PGM is still not easy to analyze for a general gauge-fixing
function $G[A_\mu,c,\bc,B]$.
Fortunately the matter becomes remarkably simple if we adopt what is
called the $\OSp(4/2)$ symmetric gauge \cite{OSpgauge} given by the
gauge-fixing function $G_{\OSp}$:
\begin{equation}
G_{\OSp}=\frac{2}{\lambda}\,\dA\!\left\{
\tr\left( A_\mu^2 + 2ic\bc\right)\right\} ,
\label{eq:GOSp}
\end{equation}
where $\lambda$ is the gauge parameter and $\dA$ is the anti-BRST
transformation:
\begin{equation}
\dA g = -g\bc,\quad \dA\bc= \Half\{\bc,\bc\},\quad
\dA c=-i\bB,\quad \dA\bB=0,\quad \left(\bB\equiv i\{c,\bc\}-B\right) .
\label{eq:dA}
\end{equation}
This is because the action of the PGM is written in a
manifestly $\OSp(4/2)$ symmetric form by introducing the superspace
$(x_\mu,\theta,\ovl{\theta})$ with Grassmann-odd coordinates $\theta$
and $\ovl{\theta}$ \cite{PGM-HK}:
\begin{equation}
S_{\OSp}= \frac{2i}{\lambda}\int\! d^4x\,
\dA\dB\left\{\tr\left(A_\mu^2+2ic\bc\right)\right\} =
\frac{2i}{\lambda}\int\! d^4x\!\int\! d\theta d\ovl{\theta}\,
\tr \left(\eta^{MN}\p_M G^\dagger\,\p_N G\right) ,
\label{eq:SOSp}
\end{equation}
where the superfield $G(x,\theta,\ovl{\theta})$ is defined by
\begin{eqnarray}
G(x,\theta,\ovl{\theta})&=& \left(1 + \ovl{\theta}\dB + \theta\dA +
\ovl{\theta}\theta\dA\dB\right)g(x) \nonumber \\
&=&g + \ovl{\theta}gc -\theta g\bc +
\ovl{\theta}\theta g\left(iB+c\bc\right) ,
\label{eq:superfieldG}
\end{eqnarray}
and the superspace metric $\eta_{MN}$
($M,N=0,1,2,3,\theta,\ovl{\theta}$) is given as
\begin{equation}
\eta_{\mu\nu}={\rm diag}\,(1,-1,-1,-1),\quad
\eta_{\ovl{\theta}\theta}=-\eta_{\theta\ovl{\theta}}=i, \quad
{\rm others}=0 .
\label{eq:etaMN}
\end{equation}
$\OSp(4/2)$ is the rotation in the superspace
$(x_\mu,\theta,\ovl{\theta})$ which leaves the metric $\eta^{MN}$
invariant. The BRST and anti-BRST transformation, $\dB$ and $\dA$,
correspond to the translation of $\ovl{\theta}$ and $\theta$,
respectively.

Then, the Parisi-Sourlas dimensional reduction mechanism
\cite{ParisiSourlas} tells that the PGM of
eq.\ (\ref{eq:SOSp}) in four dimensions is ``equivalent''
to the chiral model in two dimensions described by the action
\begin{equation}
S_{\rm chiral}=-\frac{4\pi}{\lambda}\int\! d^2 x\,\tr\left(
\p^\mu g^\dagger\,\p_\mu g\right) .
\label{eq:Schiral}
\end{equation}
The equivalence holds also in the present case of the finite
temperature system in the real-time formalism with twisted boundary
conditions $B_h$ (\ref{eq:Bh}) (the proof of ref.\ \cite{Cardy} which
does not rely on perturbation theory applies to the present case):
two spatial coordinates $(x_2,x_3)$ and two Grassmann-odd coordinates
$(\theta,\ovl{\theta})$ in the 3+1 dimensional PGM cancel to leave the
1+1 dimensional chiral model with the same boundary condition.
The equivalence implies in particular that \cite{PGM-HK}
\begin{equation}
\int\! d^4x e^{ip\cdot x}
\vev{{\rm T}\ovl{N}_\mu^a(x) A_\nu^b(0)}^{3+1}_{\beta,h}
=-2\pi\int\! d^2x e^{ip\cdot x}
\vev{{\rm T}\ovl{A}_\mu^a(x) A_\nu^b(0)}^{1+1}_{\beta,h} ,
\label{eq:NAeqAA}
\end{equation}
where the LHS (RHS) is the Green's function in the 3+1 dimensional PGM
(1+1 dimensional chiral model) with common $\beta$ and the boundary
condition $B_h$.
In \eq{eq:NAeqAA}
the four-momentum $p_\mu$ and the indices $\mu$ and $\nu$ on the LHS
should have components only in the two dimensional part $\mu,\nu=0,1$
of the RHS.
The reason why $\ovl{N}_\mu$ on the LHS of \eq{eq:NAeqAA}
is converted to $\ovl{A}_\mu\equiv A_\mu -\wt{A}_\mu$
($\wt{A}_\mu=\wt{g}\p_\mu\wt{g}$) on the RHS is the relation
$N_\mu= -i\dA\dB\left(A_\mu\right)$.
Note that $\ovl{A}_\mu$ is the Noether current of the $SU(N)_R$ symmetry
in the chiral model.

{}From eqs.\ (\ref{eq:vevinPGM2}) and (\ref{eq:NAeqAA}), we see that the
confinement condition of KO fails if the chiral $SU(N)_R$ current
$\ovl{A}_\mu$ in the 1+1 dimensional chiral model contains a massless
mode for $h$ in a finite range of $SU(N)$.
At zero temperature ($\beta=\infty$), the chiral model in 1+1
dimensions is realized in the disordered phase with a mass gap
(the system is insensitive to the boundary condition $B_h$ when
$\beta=\infty$) and hence the KO confinement condition holds
\cite{PGM-HK}.
We would like therefore to know whether the chiral model with a given
boundary condition $B_h$ undergoes a phase transition to an ordered
phase having a massless Nambu-Goldstone mode coupled to $\ovl{A}_\mu$.

\section{Analysis of the $SU(2)$ model}
\reseteqnum

Since we do not have a systematic non-perturbative technique to
analyze the $SU(N)$ chiral model in the real-time formalism with
various boundary conditions $B_h$, we shall carry out the following
approximate analysis to the $SU(2)$ chiral model.
Note that the $SU(2)$ chiral model is equivalent to the $O(4)$
non-linear $\sigma$-model via the expression
$g=\varphi_0\1 + i\sum_{a=1}^3\varphi_a\sigma^a$
($\varphi_0^2 +\varphi_a^2=1$).
Rescaling $\varphi_i$ and introducing the multiplier field $\alpha(x)$,
let us consider the following $O(4)$ non-linear $\sigma$-model system:
\begin{equation}
\calL_{O(4)}=\Half\p^\mu\vphi\cdot\p_\mu\vphi
-\Half\alpha\left(\vphi^2-\frac{1}{\lambda_0}\right) ,
\label{eq:LO(4)}
\end{equation}
where the 4 component field $\vphi=(\varphi_i)
=(\varphi_0,\ldots,\varphi_3)$ is free from the
constraint, and $\lambda_0$ is the bare coupling constant.
The boundary condition $B_h$ (\ref{eq:Bh}) for the $SU(2)$ element
$h=\exp\left(\eta_a\sigma^a/i\right)$ reads in terms of $\vphi$ as
\begin{equation}
B_h : \varphi_i(-T\!-i\beta,x)=
\left(e^{\eta_a T^a}\right)_{ij}\varphi_j(-T,x) ,
\label{eq:BCforphi}
\end{equation}
where the $4\times 4$ matrix $T^a$ ($a=1,2,3$) satisfies the property
\begin{equation}
[T^a,T^b]=2\e^{abc}T^c , \quad
\{T^a,T^b\}=-2\delta^{ab}\1 ,
\label{eq:Talgebra}
\end{equation}
and therefore we have
\begin{equation}
e^{\eta_a T^a}=\cos\eta\,\1 + \sin\eta\, T_\eta , \quad
\left(T_\eta\right)^2=-\1 , \quad
\left(\eta\equiv\left(\eta_a^2\right)^{1/2},\ \
T_\eta\equiv \eta_a T^a/\eta\right) .
\label{eq:Tproperties}
\end{equation}
The range of $\eta$ we should consider is $0\le\eta<2\pi$.

Our analysis to the model (\ref{eq:LO(4)}) is to carry out the
large-$N$ like calculation of the effective potential to determine the
phase of the model. Namely, we consider the effective potential which
is the sum of the tree part and the one-loop (trace-log) term coming
from the $\vphi$-integration.
Since the large $N$ expansion is a valid non-perturbative technique
for the (ordinary) $O(N)$ non-linear $\sigma$-model, it is expected that our
analysis here will give a qualitatively correct result for the present
$O(4)$ model with a generalized boundary condition.\footnote{
We do not know whether we can regard the present analysis as the $N=4$
case of the large $N$ expansion of an $O(N)$ model since the boundary
condition (\ref{eq:BCforphi}) is particular to $N=4$ and it breaks
explicitly the $O(3)_L$ symmetry.
}
Our question is whether the system (\ref{eq:LO(4)}) with the boundary
condition (\ref{eq:BCforphi}) undergoes a phase transition at
$\beta=\beta_c(\eta)$ from the disordered symmetric phase in the large
$\beta$ region to an ordered phase with massless Nambu-Goldstone modes
coupled to the Noether current $\ovl{A}_\mu$ (recall \eq{eq:NAeqAA}).
If at some $\beta$ the $O(4)$ model is in the Nambu-Goldstone phase
for a finite interval of the boundary condition parameter $\eta_a$,
it implies that the KO confinement condition breaks down in the
original vanishing $g_{\rm YM}$ limit of the Yang-Mills theory.

It is straightforward to apply the path-integral formalism of ref.\
\cite{NS} to the $O(4)$ model of \eq{eq:LO(4)} with unusual boundary
conditions $B_h$ (\ref{eq:BCforphi}).
We need the Green's function $D_\beta(x-y)_{ij}$ (with the $O(4)$ indices
$i,j$) on the contour of fig.\ \ref{fig:C} satisfying
\begin{equation}
\left( -\Box_C -m^2\right)D_\beta(x-y)_{ij} ,
=\delta_C(x-y)\delta_{ij} ,
\label{eq:EqforDbeta}
\end{equation}
and the boundary condition
\begin{equation}
D_\beta(\tau-i\beta,x)_{ij}=\left(e^{\eta_a T^a}\right)_{ik}
D_\beta(\tau,x)_{kj} .
\label{eq:BCforDbeta}
\end{equation}
In \eq{eq:EqforDbeta} we have $\Box_C\equiv(\p/\p\tau)^2-(\p/\p x)^2$
and $\delta_C$ is the contour $\delta$-function \cite{NS}.
$D_\beta$ is obtained in the form
\begin{equation}
D_\beta(x-y)=D_\beta^{>}(x-y)\,\theta_C(\tau_x-\tau_y)
+D_\beta^{<}(x-y)\,\theta_C(\tau_y-\tau_x) ,
\label{eq:DeqD+D}
\end{equation}
with $D_\beta^{\gtlt}$ given (in Fourier-transformed form with respect
to the spatial variable) by
\begin{eqnarray}
D_\beta^{\gtlt}(\tau;k_1)\!&=&\!\frac{1}{2i\omega f(\omega)}\Biggl\{
\left[\left(1-e^{-\beta\omega}\cos\eta\right)\1
\mp e^{-\beta\omega}\sin\eta\, T_\eta\right]e^{\mp i\omega\tau}
\nonumber\\
&&+ \left[\left(\cos\eta - e^{-\beta\omega}\right)\1
\pm \sin\eta\, T_\eta\right]e^{-\beta\omega\pm i\omega\tau}
\Biggr\} , \label{eq:Dgtlt}
\end{eqnarray}
where $\omega=|k_1|$ and
\begin{equation}
f(\omega)\equiv 1-2e^{-\beta\omega}\cos\eta+ e^{-2\beta\omega} .
\label{eq:f}
\end{equation}
For any boundary condition $B_h$, the generating functional $Z[j]$
defined by the contour $C$ of fig.\ \ref{fig:C},
\begin{equation}
Z[j]=\int_{B_h}
\calD\vphi\int_{\rm periodic}\calD\alpha\,\exp\left\{
i\int_C\left(\calL_{O(4)} + \vj\cdot\vphi + J\alpha\right)\right\} ,
\label{eq:Zj}
\end{equation}
factorizes in the limit $T\to\infty$ to the $C_1C_2$ and $C_3C_4$
parts,
\begin{equation}
Z[j]=Z[j;C_1C_2]\,Z[j;C_3C_4] .
\label{eq:ZeqZZ}
\end{equation}
We are interested in $Z[j;C_1C_2]$ with infinite real-time segments.
It is expressed as
\begin{eqnarray}
Z[j;C_1C_2]&=&\int\prod_{A=1,2}\calD\vphi_A\calD\alpha_A
\exp\Biggl\{i\int^{\infty}_{-\infty}\!dt\int\!dx\Biggl(
\Half\vphi_A\cdot\left(
D_\beta^{-1}\right)^{AB}\vphi_B \nonumber\\
&&\!\!\!-\Half\alpha_1\left(\vphi_1^2-\frac{1}{\lambda}\right)
+ \Half\alpha_2\left(\vphi_2^2-\frac{1}{\lambda}\right)
+ \vj_A\cdot\vphi_A + J_A\alpha_A\Bigg)\Bigg\} ,
\label{eq:ZjC1C2}
\end{eqnarray}
where we have defined
\begin{equation}
\vj_1(t,x)=\vj(t,x),\quad \vj_2(t,x)
=-\vj\left(t-\frac{i\beta}{2},x\right) ,
\label{eq:j1j2}
\end{equation}
and similarly for $J_A$. Note that $(\vphi_1,\vphi_2)$ corresponds to
$(\vphi,\wt{\vphi})$ in TFD.

The finite-temperature propagator $D_\beta^{AB}$ ($A,B=1,2$) is given
in momentum space as
\begin{eqnarray}
D_\beta^{AB}(k_0,k_1)&=&
\pmatrix{C & S_{-} \cr S_{+} & C\cr}
\pmatrix{
\displaystyle\frac{1}{k^2+i\epsilon} & 0 \cr
        0 & \displaystyle\frac{-1}{k^2-i\epsilon} \cr
        }
\pmatrix{C & S_{-} \cr S_{+} & C\cr} \nonumber \\
&=&\pmatrix{
\displaystyle\frac{1}{k^2+i\epsilon}-2\pi i \delta(k^2)S_{+}S_{-}&
- 2\pi i \delta(k^2) CS_{-} \cr
- 2\pi i \delta(k^2) CS_{+}&
\displaystyle\frac{-1}{k^2-i\epsilon}-2\pi i \delta(k^2)S_{+}S_{-} \cr
} ,  \label{eq:DAB}
\end{eqnarray}
where the $4\times 4$ matrices $C$ and $S_\pm$ are expressed in the
form $a\1+b T_\eta$ and should satisfy
\begin{eqnarray}
&&C^2-S_{+}S_{-}=\1 , \nonumber\\
&&S_{+}S_{-}=\frac{e^{-\beta|k_0|}}{f(|k_0|)}\left\{
\left(\cos\eta-e^{-\beta|k_0|}\right)\1
- \e(k_0)\sin\eta\, T_\eta
\right\} , \label{eq:EqsforCS}\\
&&CS_{\pm}=\pm\frac{e^{-\beta|k_0|}}{f(|k_0|)}\left\{
\left(1-e^{\mp\beta k_0}\cos\eta\right)\1
\mp e^{\mp\beta k_0}\sin\eta\, T_\eta\right\}
e^{\pm\Half\beta k_0}\e(k_0) , \nonumber
\end{eqnarray}
with $\e(x)\equiv{\rm sign}(x)$.
We present here only the explicit expression of $C$:
\begin{equation}
C=\frac{1}{\sqrt{2f(|k_0|)}}\left( u(|k_0|)\,\1 -
\frac{e^{-\beta|k_0|}\e(k_0)\sin\eta}{u(|k_0|)}T_\eta\right) ,
\label{eq:explicitC}
\end{equation}
where
$u(x)\equiv\left(1 - e^{-\beta x}\cos\eta + \sqrt{f(x)}\right)^{1/2}$.
Note that, in the particular cases of  $\eta=0$ and $\eta=\pi$, the
propagator (\ref{eq:DAB}) reduces to the familiar propagator \cite{NS}
for bosons and fermions, respectively.
The inverse propagator $\left(D_\beta^{-1}\right)^{AB}$ appearing in
\eq{eq:ZjC1C2} is given by
\begin{eqnarray}
\left(D_\beta^{-1}\right)^{AB}(k_0,k_1)&=&
\pmatrix{C & -S_{-} \cr -S_{+} & C}
\pmatrix{k^2 + i\e & 0 \cr 0 & -k^2 + i\e}
\pmatrix{C & -S_{-} \cr -S_{+} & C} \nonumber\\
&=&\pmatrix{k^2 + i\left(C^2 + S_{+}S_{-}\right)\e & -2i C S_{-}\e \cr
-2i C S_{+}\e & - k^2 + i\left(C^2 + S_{+}S_{-}\right)\e } .
\label{eq:DinvAB}
\end{eqnarray}
Although the boundary conditions $B_h$ breaks explicitly
the $O(3)_L$ symmetry except the cases $\eta=0$ and $\pi$, the effect of
the breaking in the inverse propagator appears only at the $i\e$ parts.

As stated at the beginning of this section, we consider the
effective potential $V_\beta(\vphi_A,\alpha_A)$ which is the sum of
the tree part and the trace-log term coming from the $\vphi$-integration:
\begin{eqnarray}
V_\beta(\vphi_A,\alpha_A)&=&
\Half\alpha_1\left(\vphi_1^2-\frac{1}{\lambda_0}\right)
-\Half\alpha_2\left(\vphi_2^2-\frac{1}{\lambda_0}\right)
\nonumber\\
&&-i\frac{1}{2}\int\!\frac{d^2k}{(2\pi)^2}\,\tr\ln
\left[D_\beta^{-1}
- \pmatrix{\alpha_1\1&0\cr 0&-\alpha_2\1\cr}\right] .
\label{eq:Vbeta}
\end{eqnarray}
The phase of the theory is determined by the stationary condition of
$V_\beta(\vphi_A,\alpha_A)$ with respect to $\vphi_A$ and
$\alpha_A$ ($A=1,2$). The stationary points exist in the subspace
$\alpha_1=\alpha_2$ ($V_\beta$ (\ref{eq:Vbeta}) is singular
when $\alpha_1\ne\alpha_2$ \cite{NS}).
Therefore, the conditions to determine the groundstate are
\begin{eqnarray}
&&\frac{\p}{\p\vphi_A}V_\beta\Bigg\vert_{\alpha_1=\alpha_2=\alpha}
= \alpha\vphi_A=0 ,
\label{eq:dVdvphi}\\
(-)^{A+1}\kern-20pt
&&\frac{\p}{\p\alpha_A}V_\beta\Bigg\vert_{\alpha_1=\alpha_2=\alpha}
= \Half\left(\vphi_A^2-\frac{1}{\lambda_0}\right)
+\frac{N}{8\pi}\int^{\Lambda}_{-\Lambda}\frac{dk}{\sqrt{k^2+\alpha}}
\,\frac{1 -  e^{-2\beta\sqrt{k^2+\alpha}}}
{f\!\left(\sqrt{k^2+\alpha}\right)}=0 ,
\label{eq:dVdalpha}
\end{eqnarray}
where $N=\tr \1=4$ in the present case, and we have introduced the cut-off
$\Lambda$ for the $k$ (spatial momentum) integration.
The first condition (\ref{eq:dVdvphi}) tells that we have either
$\alpha=0$ or $\vphi_A=0$, and from the second condition
(\ref{eq:dVdalpha}) we see that $\vphi_1^2=\vphi_2^2$.
The second condition (\ref{eq:dVdalpha}) is rewritten using the mass
gap $m$ at zero temperature instead of $\lambda_0$ and $\Lambda$.
Note that $m$ is determined by \eq{eq:dVdalpha} with $\beta=\infty$,
$\alpha=m^2$ and $\vphi_A=0$ as
\begin{equation}
\frac{2\pi}{N\lambda_0}=\int^{\Lambda}_0\frac{dk}{\sqrt{k^2+m^2}}
=\ln\!\left(\frac{2\Lambda}{m}\right) ,
\label{eq:m}
\end{equation}
which implies a familiar formula
$m=2\Lambda\exp\left(-2\pi/N\lambda_0\right)$.
Using \eq{eq:m} we have
\begin{equation}
(-)^{A+1}
\frac{\p}{\p\alpha_A}V_\beta\Bigg\vert_{\alpha_1=\alpha_2=\alpha}
= \Half\vphi_A^2 +\frac{N}{4\pi}K_\beta(\alpha)=0 ,
\label{eq:dVdalpha-m}
\end{equation}
with $K_\beta(\alpha)$ defined by
\begin{equation}
K_\beta(\alpha)\equiv
\int^{\infty}_0\! dk\left(\frac{1}{\sqrt{k^2+\alpha}}
\,\,\frac{1 -  e^{-2\beta\sqrt{k^2+\alpha}}}
{f\!\left(\sqrt{k^2+\alpha}\right)}
-\frac{1}{\sqrt{k^2+m^2}}\right) .
\label{eq:K}
\end{equation}
The property of our $O(4)$ model is qualitatively different between
the periodic boundary condition ($\eta=0$) and the twisted one
($0<\eta<2\pi$). We shall discuss the two cases separately below.

\noindent
\underline{Periodic boundary condition}

It is easily seen that \eq{eq:dVdalpha-m} with $\eta=0$ has no
solution with $\alpha=0$ for any $\beta$ since $K_\beta(\alpha=0)$ is
positive and infrared divergent (note that $f(x)=(1-e^{-x})^2$ when
$\eta=0$). Therefore we have always $\vphi_A=0$.
The expectation value of $\alpha$, which is the $(\mbox{\rm mass})^2$
of the $\vphi$ particles, is determined by \eq{eq:dVdalpha-m} with
$\eta=0$ and $\vphi^2=0$, and it is a monotonically increasing function
of the temperature $1/\beta$. Therefore, the system with periodic
boundary condition is in the disordered phase for any $\beta$.

\noindent
\underline{Twisted boundary condition}

In this case, there is a critical temperature $\beta_c(\eta)$
determined by the condition $K_{\beta=\beta_c(\eta)}(0)=0$ or
explicitly,
\begin{equation}
\ln\!\left(\frac{m\beta_c(\eta)}{2}\right)=
\int_0^{\infty}\!dx\,\ln x\cdot \frac{d}{dx}\!\left(
\frac{\cosh(x/2)\sinh(x/2)}{\sinh^2(x/2)+\sin^2(\eta/2)}
\right) .
\label{eq:Eqforbetac}
\end{equation}
The critical temperature $\beta_c(\eta)$ as a function of the boundary
condition parameter $\eta$ ($0\le\eta<2\pi$) is depicted in
fig.\ \ref{fig:betac} in units of $\beta_c(\pi)=\pi e^{-\gamma}/m$
($\beta_c$ for the anti-periodic boundary condition
\cite{DolanJackiw,GN-model}).
As seen from fig.\ \ref{fig:betac}, $\beta_c(\eta)$ is a monotonically
increasing function of $\sin(\eta/2)$ and vanishes only when $\eta=0$.
The meaning of $\beta_c(\eta)$ is that $K_\beta(0)>0$ ($<0$) when
$\beta>\beta_c$ ($\beta<\beta_c$) (see fig.\ \ref{fig:k}).
The two phases separated by $\beta_c(\eta)$ are as
follows:

\epsfhako{betac.eps}{htb}{10cm}{
The critical temperature $\beta_c(\eta)$ in units of
$\beta_c(\pi)$.}{fig:betac}{\vspace{-.5cm}}

\epsfhako{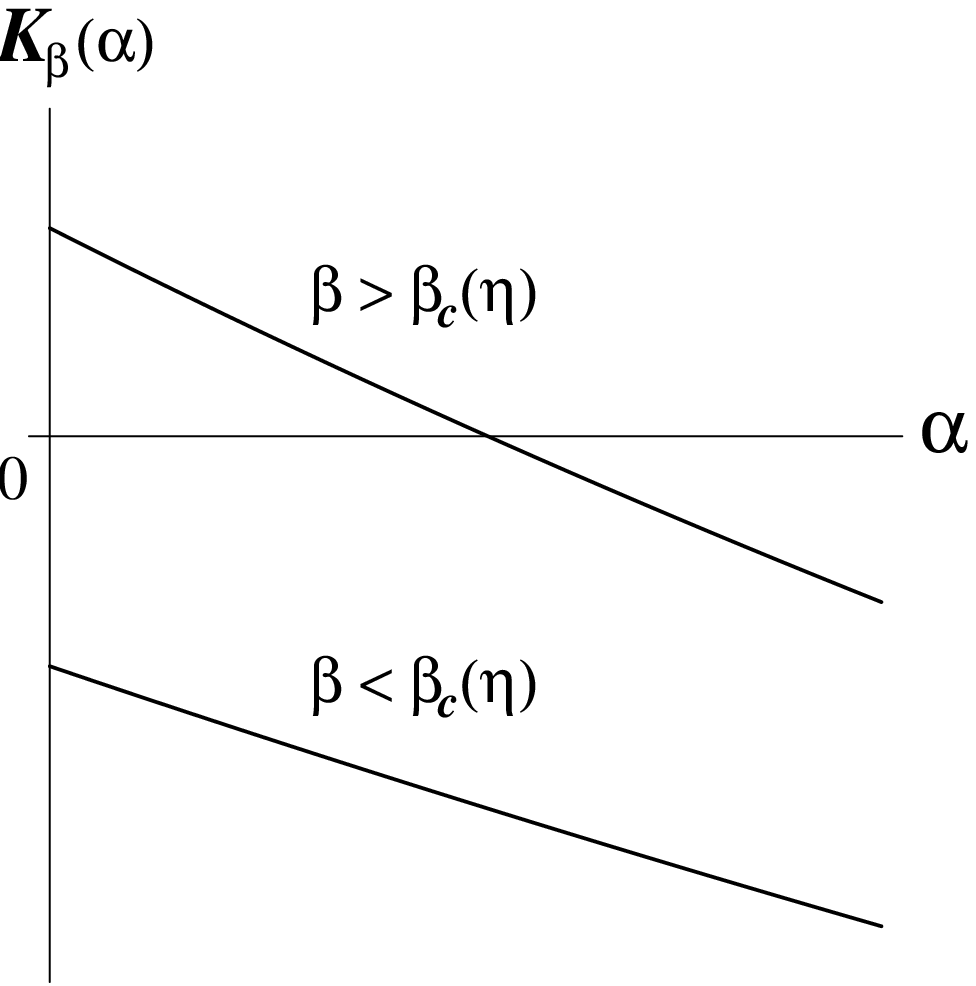}{htb}{8cm}{The function $K_\beta(\alpha)$ with $\beta$
below and above $\beta_c(\eta)$.}{fig:k}{}

\noindent
i) Low temperature region $\beta>\beta_c(\eta)$:
Eq.\ (\ref{eq:dVdalpha-m}) has no solution of the type $(\vphi_A\ne
0,\alpha=0)$ since we have $K_\beta(\alpha=0)>0$ when
$\beta>\beta_c$ (see fig.\ \ref{fig:k}). Therefore, the system is in
the disordered phase. The non-vanishing value of $\alpha$ determined
by \eq{eq:dVdalpha-m} with $\vphi_A=0$, i.e., the intercept of the curve
of fig.\ \ref{fig:k} with the $\alpha$-axis, gives the
$(\mbox{mass})^2$ of the $\vphi$ excitation.

\noindent
ii) High temperature region $\beta<\beta_c(\eta)$:
Eq.\ (\ref{eq:dVdalpha-m}) has no solution of
the type $(\vphi_A=0, \alpha>0)$ since $K_\beta(\alpha)$, which is
a monotonically decreasing function of $\alpha$, is negative definite
when $\alpha>0$ (see fig.\ \ref{fig:k}).\footnote{
The $k$-integration in \eq{eq:K} can be continued to the
$\alpha<0$ region, and \eq{eq:dVdalpha-m} with $\vphi_A=0$ has a
negative $\alpha$ solution even when $\beta<\beta_c$.
However, we do not adopt this solution since it implies that the
$\vphi$ excitation is tachyonic.
}
Therefore the system is realized in the ordered phase with
$(\vphi_A\ne 0, \alpha=0)$.
The Noether current of $SU(2)_R$,
$\ovl{A}^a_\mu=(A^a_\mu)_{A=1}-(A^a_\mu)_{A=2}$
($A^a_\mu=\e^{abc}\varphi_b\p_\mu\varphi_c+\varphi_0{\lrp}_\mu\varphi_a$),
is coupled to the Nambu-Goldstone mode irrespectively of the direction
of the expectation value $\vphi_A$.

The above results are summarized as follows.
When $\beta>\beta_c(\pi)$, our $O(4)$ model is in the disordered phase
for all the boundary conditions $B_h$. When $\beta<\beta_c(\pi)$, the
models with $\eta$ in the range
$\sin\left(\beta_c^{-1}(\beta)\right)<\sin(\eta/2)\le 1$
($\beta_c^{-1}$ is the inverse function of $\beta_c$)
are in the Nambu-Goldstone phase, while the models in the other range
of boundary conditions are in the disordered phase.
The range of the boundary conditions corresponding to the ordered
phase increases as we increase the temperature $1/\beta$.
Translating this result back to the original $SU(2)$ PGM in 3+1
dimensions obtained as the vanishing $g_{\rm YM}$ limit of Yang-Mills
theory, the KO confinement condition is satisfied at
low temperature $\beta>\beta_c(\pi)$, however it breaks down at high
temperature $\beta<\beta_c(\pi)$ due to the contribution of the
twisted boundary condition sectors in the average (\ref{eq:vevinPGM2})
with $\calO=\ovl{N}_\mu A_\nu$. The Green's function (\ref{eq:limNA})
develops a massless pole when $\beta<\beta_c(\pi)$.

Some comments are in order.
First, we should comment on the compatibility of our result (i.e.,
that the Nambu-Goldstone phase is realized when $\beta<\beta_c(\eta)$
for the twisted sectors) with Coleman's theorem which
forbids the Nambu-Goldstone bosons in two dimensions.
The origin of the absence of the Nambu-Goldstone bosons is that the
(ordinary) massless scalar propagator in 1+1 dimensions
\begin{equation}
\int\! \frac{d^2k}{(2\pi)^2}\,\frac{e^{ik\cdot x}}{k^2+i\e} ,
\label{eq:masslesspropagator}
\end{equation}
does not exist because of infrared divergence.
In the present case of the $O(4)$ model with a twisted boundary
condition, the massless propagator in momentum space (see \eq{eq:DAB})
takes in the infrared $k_\mu\sim 0$ the following form:
\begin{equation}
D_\beta^{AB}(k)\sim
\pmatrix{
\displaystyle {\cal P}\!\left(\frac{1}{k^2}\right)
+i\pi\e(k_0)\delta(k^2)\cot\!\left(\frac{\eta}{2}\right) T_\eta
&\displaystyle
i\pi\left(\1+\cot\!\left(\frac{\eta}{2}\right)T_\eta\right)
\epsilon(k_0)\delta(k^2) \cr
\displaystyle
-i\pi\left(\1-\cot\!\left(\frac{\eta}{2}\right)T_\eta\right)
\epsilon(k_0)\delta(k^2)
&\displaystyle -{\cal P}\!\left(\frac{1}{k^2}\right)
+i\pi\e(k_0)\delta(k^2)\cot\!\left(\frac{\eta}{2}\right) T_\eta
} ,
\label{eq:D-IR}
\end{equation}
where ${\cal P}$ denotes the principal part.
Coleman's theorem is evaded since the Fourier transform of the RHS of
\eq{eq:D-IR} does exist:
\begin{equation}
\int\! \frac{d^2k}{(2\pi)^2}\left(
{\cal P}\!\left(\frac{1}{k^2}\right),\, i\pi\e(k_0)\delta(k^2)
\right)e^{ik\cdot x}= \left(
-\Half\theta(x^2),\, \frac{1}{4}\e(x_0)\theta(x^2) \right) ,
\label{eq:FTs}
\end{equation}
where $\theta(x)$ is the step function, $\theta(x)=(\e(x) +1)/2$.
An intuitive reason why the deconfining transition occurs in our model
is that the infrared singularity (in the perturbative ordered phase)
which caused the disordered phase at zero temperature is weakened by
the twisted boundary conditions.
The effect of the boundary condition becomes stronger as we raise the
temperature and hence triggers the transition to the ordered phase.

Our second comment is on the imaginary-time formalism.
In this paper we have employed the real-time formalism of finite
temperature field theory since our interest is in the KO confinement
condition, which needs continuous four-momentum and cannot be
discussed in the imaginary-time formalism.
Forgetting this fact for the moment, let us consider what
happens if we adopt the imaginary-time formalism defined by the
straight vertical time contour $[0,-i\beta]$ (cf.\ fig.\ \ref{fig:C})
in the above analysis of the $O(4)$ non-linear $\sigma$-model.
Then, for the exactly periodic sector with $\eta=0$ we get the same
conclusion that $\bvev{\alpha}>0$ and $\bvev{\vphi}=0$ for all
$\beta$.
For the twisted boundary condition sector, however, we get a
completely different result from the real-time formalism: the
disordered phase with $\bvev{\vphi}=0$ is realized for all $\beta$.
This is because in the imaginary-time formalism with a twisted
boundary condition, $\vphi$ has no zero-mode to develop an
expectation value.
In particular, expanding $\vphi$ into modes,
$\vphi(\tau,\bm{x})=\sum_{n=-\infty}^{\infty}\vphi_n(\bm{x})
e^{i(2n+1)\pi\tau/\beta}$ ($0\!\le\!\tau\!\le\!\beta$) for the
anti-periodic boundary condition sector ($\eta=\pi$),
the large $N$ analysis shows that no modes $\vphi_n$ can develop
expectation values. The expectation value $\bvev{\alpha}$ is a
monotonically decreasing function of the temperature  $1/\beta$ and
becomes negative for $\beta<\beta_c(\pi)$.
Note that a negative $\bvev{\alpha}$ is not a trouble in this case
since the effective $(\mbox{mass gap})^2$ is given by
$\bvev{\alpha}+(\pi/\beta)^2$, which is seen to be always positive.
The discrepancy between real-time and imaginary-time formalisms in the
case of twisted boundary conditions may seem strange, but it is
not a problem since the twisted sector is not an ordinary
statistical mechanics system. This discrepancy, however, is
disappointing for an attempt to analyze the original chiral model by
the Monte Carlo simulation since it is possible only in the
imaginary-time formalism.

\section{Summary and discussion}
\reseteqnum

In this paper, as a first step toward the understanding of the
deconfining transition in Yang-Mills theory in the sense of the
breakdown of the color confinement condition of Kugo and Ojima, we
have studied the model obtained by taking the limit of vanishing gauge
coupling constant. This model at zero temperature is known to satisfy
the KO confinement condition.
Adopting a special gauge-fixing function, the system in 3+1
dimensions in the real-time formalism is reduced to a ``sum'' of
chiral models in 1+1 dimensions with various boundary conditions
concerning the time contour.
As a qualitative approximation to the $SU(2)$ chiral model we analyzed
the equivalent $O(4)$ non-linear $\sigma$-model using the large-$N$ like
analysis. We found that the $O(4)$ model with a twisted boundary
condition undergoes a transition to Nambu-Goldstone phase because the
infrared singularity is softened by the boundary condition.
This implies the breakdown of the KO confinement condition in the 3+1
dimensional model.

The pure-gauge model we considered in this paper, namely the vanishing
$g_{\rm YM}$ limit of the Yang-Mills theory, is physically trivial,
and the transition we have found is not the singularity of the free
energy for the 3+1 dimensional pure-gauge model.
However, since the confinement mechanism of KO has an intimate
relationship with the gauge field disorder in the direction of local
gauge transformation \cite{Hata-restoration}, the breakdown of the
KO confinement condition we observed should suggest the deconfining
transition in the real Yang-Mills theory.\footnote{
See ref.\ \cite{Izawa} for an attempt to show the confinement by the
KO mechanism in the real Yang-Mills theory on the basis of the
pure-gauge model.}

\vskip1cm
\centerline{\large\bf Acknowledgements}

The authors would like to thank K.~I.~Izawa and M.~G.~Mitchard
for valuable discussions.

%%%%%%%%%%%%%%%%%%%%%%%%%%%%%%%%%%%%%%%%%%%%%%%%%%%%%%%%%
\newcommand{\J}[4]{{\sl #1} {\bf #2} (19#3) #4}
\newcommand{\MPL}{Mod.\ Phys.\ Lett.}
\newcommand{\NP}{Nucl.\ Phys.}
\newcommand{\PL}{Phys.\ Lett.}
\newcommand{\PR}{Phys.\ Rev.}
\newcommand{\PRL}{Phys.\ Rev.\ Lett.}
\newcommand{\AP}{Ann.\ Phys.}
\newcommand{\CMP}{Commun.\ Math.\ Phys.}
\newcommand{\PTP}{Prog.\ Theor.\ Phys.}
%%%%%%%%%%%%%%%%%%%%%%%%%%%%%%%%%%%%%%%%%%%%%%%%%%%%%%%%%

%%%%%%%%%%%%%%%%%%%%%%%%%%%%%%%%%%%%%%%%%%%%%%%%%%%%%%%%%


\begin{thebibliography}{99}

\bibitem{Polyakov} A.\ M.\ Polyakov, \J{\PL}{72B}{78}{447}. % 180

\bibitem{Susskind} L.\ Susskind, \J{\PR}{D20}{79}{2610}. %180

\bibitem{KO} T.\ Kugo and I.\ Ojima, \J{\PTP\ Suppl.}{66}{79}{1}. % 187

\bibitem{HataNiigata} H.\ Hata and I.\ Niigata,
\J{\NP}{B389}{93}{133}. %250

\bibitem{FiniteT-HK} H.\ Hata and T.\ Kugo, \J{\PR}{D21}{80}{3333}.%262

\bibitem{NS} A.\ J.\ Niemi and G.\ W.\ Semenoff,
\J{\AP}{152}{84}{105}. % 292

\bibitem{TakaUme} Y.\ Takahashi and H.\ Umezawa,
\J{Collective Phenomena}{2}{75}{55}. % 293

\bibitem{UMT} H.\ Umezawa, H.\ Matsumoto and M.\ Tachiki,
``Thermo Field Dynamics and Condensed States'' (North-Holland,
Amsterdam, 1982). % 293

\bibitem{Ojima} I.\ Ojima, \J{\AP}{137}{81}{1}. % 345

\bibitem{PGM-H} H.\ Hata, \J{\PL}{B143}{84}{171}. % 409

\bibitem{PGM-HK} H.\ Hata and T.\ Kugo, \J{\PR}{D32}{85}{938}.%409

\bibitem{ParisiSourlas} G.\ Parisi and N.\ Sourlas,
\J{\PRL}{43}{79}{744}. % 428

\bibitem{PolyakovWiegmann} A.\ Polyakov and P.\ B.\ Wiegmann,
\J{\PL}{B131}{83}{121};\\
P.\ B.\ Wiegmann, \J{\PL}{B141}{84}{217}; \J{}{B142}{84}{173}. %430

\bibitem{Hata-restoration} H.\ Hata,
\J{\PTP}{67}{82}{1607};\J{}{69}{83}{1524}. %438

\bibitem{OSpgauge} R.\ Delbourgo and P.\ D.\ Jarvis,
\J{J.\ Phys.}{A15}{82}{611} % 599

\bibitem{Cardy} J.\ L.\ Cardy, \J{\PL}{125B}{83}{470}. % 657

\bibitem{DolanJackiw} L.\ Dolan and R.\ Jackiw,
\J{\PR}{D9}{74}{3320}. %998

\bibitem{GN-model} L.\ Jacobs, \J{\PR}{D10}{74}{3956};\\
B.\ Harrington and A.\ Yildiz, \J{\PR}{D11}{75}{1499}. % 998

\bibitem{Izawa} K.\ I.\ Izawa, \J{\PTP}{90}{93}{911}. %1193

\end{thebibliography}
\end{document}